\documentclass[twocolumn,
prd,amssymb,amsmath,preprintnumbers,
floatfix,aps,nofootinbib]{revtex4-1}

\usepackage{amsmath}
\usepackage{amssymb}
\usepackage{graphicx}
\usepackage{mathrsfs}
\usepackage{cancel}
\usepackage{soul}
\usepackage{color}
\usepackage{multirow}
\usepackage{bm}

\newcommand{\beq}{\begin{equation}}
\newcommand{\eeq}{\end{equation}}
\newcommand{\bea}{\begin{eqnarray}}
\newcommand{\eea}{\end{eqnarray}}

\newcommand{\black}{\color{black}}

\allowdisplaybreaks[1]

\def\oRe{\operatorname{Re}}
\def\oIm{\operatorname{Im}}

\def\spa#1.#2{\left\langle#1\,#2\right\rangle}
\def\spb#1.#2{\left[#1\,#2\right]}

\begin{document}

\title{ Probe $CP$-violating $H\gamma\gamma$ coupling through interferometry}

\author{Xia Wan}
\email{wanxia@snnu.edu.cn}
\affiliation{School of Physics $\&$ Information Technology, Shaanxi Normal University, Xi'an 710119, China}

\author{You-Kai Wang}
\email{wangyk@snnu.edu.cn}
\affiliation{School of Physics $\&$ Information Technology, Shaanxi Normal University, Xi'an 710119, China}

\date{\today}

\begin{abstract}

The diphoton invariant mass distribution of interference between
 $gg\to H \to \gamma\gamma$ and $gg\to \gamma\gamma$
is almost antisymmetric around the Higgs mass $M_H$. We propose
a new observable $A_{\text{int}}$ to quantify this effect, which is a ratio of
a sign-reversed integral around $M_H$ ( e.g. $\int^{M_H}_{M_H-5~\mbox{GeV}}
-\int_{M_H}^{M_H+5~\mbox{GeV}}$) and the cross section of the
Higgs signal.
We study $A_{\text{int}}$ both in Standard Model (SM) and new physics with various $CP$-violating
$H\gamma\gamma$ couplings.
The $A_{\text{int}}$ in SM
 could reach a value of $10\%$,
while for $CP$-violating $H\gamma\gamma$ couplings
$A_{\text{int}}$ could range from $10\%$ to $-10\%$,
which is probable to be detected in HL-LHC experiment.
The $A_{\text{int}}$ with both $CP$-violating $H\gamma\gamma$ and $Hgg$ couplings are
also studied and its value range is further extended.

\end{abstract}

\maketitle

\section{Introduction\label{Introduction}}

The $CP$ violation, as one of the three Sakharov conditions~\cite{Sakharov:1967dj},
is necessary when explaining the matter-antimatter
 asymmetry in our universe~\cite{Ade:2013zuv}.
Its source could have close relation with Higgs dynamics~\cite{Accomando:2006ga,Morrissey:2012db}.
Thus the $CP$ properties of the 125~GeV Higgs boson with spin zero is proposed to
be probed in various channels at the Large Hadron Collider (LHC)~\cite{Gao:2010qx,Bolognesi:2012mm,Gritsan:2016hjl,Ellis:2013lra,Buckley:2015vsa,Li:2015kxc,Hayreter:2016aex,Hagiwara:2016zqz,Dolan:2014upa,Chen:2014ona,Korchin:2014kha,Chen:2017plj,Bian:2017jpt,Brehmer:2017lrt,Khachatryan:2014kca,Khachatryan:2016tnr,Sirunyan:2017tqd,CMS:2018bwq,Aad:2016nal,Aaboud:2017vzb}.
Among them, the golden channel $H\to ZZ\to 4\ell$ has been studied extensively
and it gives relative stringent experimental constraints~\cite{Khachatryan:2014kca,Sirunyan:2017tqd,CMS:2018bwq,Aaboud:2017vzb}.
By contrary, the $H\to\gamma\gamma$ process
 is another golden channel to discover
Higgs boson and has a relative clean signature, but
it suffers from lacking of $CP$-odd observable constructed from
the self-conjugated diphoton signal kinematic variables.
The $CP$ property in $H\gamma\gamma$ coupling could also be studied in
$H\to\gamma^\ast\gamma^\ast\to 4\ell$ process \cite{Voloshin:2012tv,Bishara:2013vya,Chen:2014gka}, however, it is challenged by the low conversion rate
 of the off-shell photon decaying into two leptons.
In this paper, we study the $CP$ property in $H\gamma\gamma$ coupling
through the interference between $gg\to H \to \gamma\gamma$ and $gg\to \gamma\gamma$.

This interference has been studied in many papers~\cite{Dicus:1987fk,
Dixon:2003yb,Martin:2012xc,deFlorian:2013psa,Martin:2013ula,Dixon:2013haa,Campbell:2017rke,Djouadi:2016ack}.
Compared to the Breit-Wigner lineshape of the Higgs boson's signal,
 the lineshape of the interference term could be roughly divided into two parts:
 one is symmetric around $M_H$ and the other is antisymmetric around $M_H$.
These two kinds of interference lineshapes
have different effects:
after integrating over a symmetric mass region around $M_H$, the symmetric interference lineshape could reduce the signal Breit-Wigner cross section by $\sim2\%$~\cite{Campbell:2017rke}; while the antisymmetric one has no contribution to the total cross section, but could
 distort the signal lineshape, and shift the resonance mass peak by
$\sim150$~MeV~\cite{Martin:2012xc,Dixon:2013haa}.
Besides,
a variable $A_i$
is proposed~\cite{Lillie:2007ve, Bian:2015hda} to quantify the interference effect in a sophisticated way,
which defines a sign-reversed integral around $M_H$ (e.g. $\int_{M_H-5~\textup{GeV}}^{M_H}dM
-\int^{M_H+5~\textup{GeV}}_{M_H}dM$)
in its numerator and a sign-conserved integral around $M_H$ (e.g. $\int_{M_H-5~\textup{GeV}}^{M_H}dM
+\int^{M_H+5~\textup{GeV}}_{M_H}dM$) in its denominator,
with both the integrands being overall lineshape, which is superimposed by the signal lineshape, the symmetric interference lineshape and the antisymmetric interference lineshape.
In principal, all three
effects from interference, changing signal cross section, shifting resonance mass peak and
$A_i$ (the ratio of sign-reversed integral and sign-conserved integral),
could be used to probe $CP$ violation in $H\gamma\gamma$
coupling, but their sensitivities are different.
As the symmetric interference lineshape is contributed
mainly from the Next-to-Leading order
 while the antisymmetric one from the leading order~\cite{Dixon:2003yb,Campbell:2017rke},
 the effect from antisymmetric interference lineshape has
better sensitivity,
which means the latter two effects could be more sensitive to $CP$ violation.

Nevertheless, experimentally $A_i$ is not trivial, and could be affected a lot by the mass
uncertainty of $M_H$~\cite{Bian:2015hda}. The main reason is once $M_H$ was deviated a little,
the sign-reversed integral in the numerator would get a large extra value from the signal lineshape.
To solve this problem, we suggest to separate the antisymmetric interference lineshape from the overall
lineshape firstly, then replace the integrand in the numerator with only the antisymmetric interference lineshape.
Thus the effect from the mass uncertainty is suppressed in the observable. The new modified observable
is named as $A_{\text{int}}$ and it is used to quantify the interference effect in our analysis.

In this paper,
we study to probe the $CP$ property in $H\gamma\gamma$ coupling
through
interference between
$gg\to H \to \gamma\gamma$ and $gg\to \gamma\gamma$.
The rest of the paper is organized as follows. In Section~\ref{section:calc},
we introduce an effective model with a $CP$-violating $H\gamma\gamma$ coupling,
 and calculate the interference between $gg\to H \to\gamma\gamma$ and $gg\to \gamma\gamma$.
 Then we introduce the observable $A_{\text{int}}$, and study its dependence on $CP$ violation.
In Section~\ref{section:simulation}, we simulate the lineshapes of the signal and the interference,
 and get the $A_{\text{int}}$ in SM and various $CP$-violation cases.
 After that, we estimate the feasibility to measure $A_{\text{int}}$ at LHC and High Luminosity Large Hadron Collider(HL-LHC).
In Section~\ref{section:hgg}, we build a general framework with both $CP$
violating $H\gamma\gamma$ and $Hgg$ couplings and study the $A_{\text{int}}$ in a same procedure
as above.
In Section~\ref{section:conclusion}, we give a conclusion and discussion.

\section{Theoretical calculation\label{section:calc}}

The effective model with a $CP$-violating $H\gamma\gamma$ coupling is given as,

\bea
        \mathcal{L}_{\rm h} &=&
\frac{c_\gamma \cos\xi_\gamma}{v}~h\,F_{\mu\nu} F^{\mu\nu} + \frac{c_\gamma \sin\xi_\gamma}{2v}~h\,F_{\mu\nu}\tilde{F}^{\mu\nu} \nonumber \\
&&+ \frac{c_g}{v}~h\,G^a_{\mu\nu}G^{a\mu\nu},
\label{eqn:HEF}
\eea
where $F$, $G^a$ denote the $\gamma$ and gluon field strengths, $a = 1,...,8$ are $SU(3)_c$ adjoint representation indices for the gluons, $v = 246$~GeV is the electroweak vacuum expectation value, the dual field strength is defined as $\tilde{X}^{\mu\nu}=\epsilon^{\mu\nu\sigma\rho}X_{\sigma\rho}$, $c_\gamma$ and $c_g$ are effective couplings in SM at leading order, $\xi_\gamma\in[0,2\pi)$ is a phase that parametrize $CP$ violation.
When $\xi_\gamma=0$, it is the SM case; when $\xi_\gamma\ne0$, there must exist $CP$ violation
(except for $\xi_\gamma=\pi$) and new physics beyond SM.
This kind of parametrization could make sure that the total signal
strength of Higgs decaying to diphoton is equal to one as predicted by SM.

In SM at leading order, $c_\gamma$ is introduced by fermion and $W$ loops and $c_g$ is introduced by fermion loops only, which have the expressions as
\bea
&&c_g =\frac{\alpha_s}{16\pi }\sum_{f=t,b}F_{1/2}(4m^2_f/\hat{s}), \\
&&c_\gamma =\frac{\alpha}{8\pi }\left[F_1(4m^2_W/\hat{s})
+\sum_{f=t,b}N_c Q^2_f F_{1/2}(4m^2_f/\hat{s}) \right ]~.
\eea
where $\alpha_s(\alpha)$ are running QCD(QED) couplings, $N_c=3$, $Q_f$ and $m_f$ are electric charge and mass of fermions, and
\bea
F_{1/2}(\tau) = -2\tau[1+(1-\tau)f(\tau)], \\
F_1(\tau) = 2+3\tau[1+(2-\tau)f(\tau)],
\eea
\beq
f(\tau) = \left\{
\begin{array}{ll}
{\rm arcsin}^2 \sqrt{1/\tau} & \tau \geq 1~, \\
-\frac{1}{4} \left[ \log \frac{1 + \sqrt{1-\tau } }
{1 - \sqrt{1-\tau} } - i \pi \right]^2 \ \ \ & \tau <1~.
\end{array} \right.
\eeq

The helicity amplitudes for $gg\to H\to\gamma\gamma$ and  $gg\to \gamma\gamma$
can be written as~\cite{Martin:2012xc,Bern:2001df,Bern:2002jx},
\bea
\mathcal{M}
&=& -e^{-i h_3\xi_\gamma}\delta^{h_1h_2}\delta^{h_3h_4}\delta^{a b}\frac{M^4_{\gamma\gamma}}{v^2}
\frac{4c_gc_\gamma}{M^2_{\gamma\gamma}-M^2_H+iM_H\Gamma_H}  \nonumber \\
&& + 4\alpha\alpha_s\delta^{a b}\sum_{f=u,d,c,s,b} Q^2_f
\mathcal{A}_{\text{box}}^{h_1 h_2 h_3h_4}~,
\label{eqn:amp}
\eea
where $a,b$ are the same as $a$ in Eq.~\eqref{eqn:HEF},
the spinor phases (see their concrete formulas in~\cite{Bern:2001df,Bern:2002jx} and~\cite{Chen:2017plj}) are dropped for simplicity,
 $h_i$s are helicities of outgoing gluons and photons,
$Q_f$ is the electric charge of fermion,
$\mathcal{A}_{\text{box}}^{h_1 h_2 h_3h_4}$ are reduced 1-loop helicity amplitudes of
$gg\to \gamma\gamma$ mediated by
 five flavor quarks, while the contribution from top quark is
much suppressed~\cite{Dicus:1987fk} and is neglected in our analysis.
The $\mathcal{A}_{\text{box}}$ for non-zero interference are~\cite{Martin:2012xc,Bern:2001df,Bern:2002jx}
\bea
&&\mathcal{A}_{\text{box}}^{++++}=\mathcal{A}_{\text{box}}^{----}=1, \nonumber \\
&&\mathcal{A}_{\text{box}}^{++--}=\mathcal{A}_{\text{box}}^{--++}=  \nonumber \\
&&-1 + z \ln\left (\frac{1 + z}{1 - z}\right) - \frac{1 + z^2}{4}\left[\ln^2
\left(\frac{1 + z}{1 - z}\right)+\pi^2\right],
\label{eqn:appmm}
\eea
where $z=\cos\theta$, with $\theta$ being the scattering angle of $\gamma$ in
diphoton center of mass frame. It maybe noticed by a careful reader that we use
the formulas for $\mathcal{A}_{\text{box}}^{++++/----}$ and $\mathcal{A}_{\text{box}}^{++--/--++}$ as same as in \cite{Bern:2001df,Bern:2002jx}, which are swapped in \cite{Martin:2012xc}. That is because the convention we used here are for outgoing gluons, the
helicities would be changed to a reversed sign if for incoming gluons.
It is also worthwhile to notice that Eq.~\eqref{eqn:amp} is different from Eq.~(2) in Ref.~\cite{Martin:2012xc} because of the $e^{-i h_3\xi_\gamma}$ factor, which determines that the Higgs signal strength are not affected by the $CP$-violation factor $\xi_\gamma$, but
the interference strength has a simple $\cos\xi_\gamma$ dependence (see Eqs.~\eqref{eqn:sigmasig}\eqref{eqn:sigmaint}).

After considering interference, the lineshape over the smooth background is composed
of both lineshapes of signal and interference, which can be expressed by

\bea
\frac{d\sigma_{\text{sig}}}{dM_{\gamma\gamma}}
&=&\frac{G(M_{\gamma\gamma})}{128\pi M_{\gamma\gamma}}
\frac{ |c_g c_\gamma|^2 }
{(M^2_{\gamma\gamma}-M^2_H)^2+M^2_H\Gamma^2_H}\times \int dz,
\label{eqn:sigmasig}
 \\
\frac{d\sigma_{\text{int}}}{dM_{\gamma\gamma}}
&=&\frac{G(M_{\gamma\gamma})}{128\pi M_{\gamma\gamma}}
\frac{(M^2_{\gamma\gamma}-M^2_H)\oRe\left(c_g c_\gamma \right)
+M_H\Gamma_H \oIm\left(c_g c_\gamma \right)
}
{(M^2_{\gamma\gamma}-M^2_H)^2+M^2_H\Gamma^2_H}
\nonumber\\
&&\times\int dz
[\mathcal{A}_{\text{box}}^{++++}+\mathcal{A}_{\text{box}}^{++--}]\times\cos\xi_\gamma,
\label{eqn:sigmaint}
\eea
where $\sigma_{\text{sig}}, \sigma_{\text{int}}$ are cross sections from signal term and interference term
respectively, $M_{\gamma\gamma}=\sqrt{\hat{s}}$, the integral region of $z$ depends on the detector angle coverage in experiment,
$G(M_{\gamma\gamma})$ is gluon-gluon luminosity function written as
\beq
G(M_{\gamma\gamma})=\int^1_{M^2_{\gamma\gamma}/s}\frac{dx}{sx}[g(x)g(M^2_{\gamma\gamma}/(sx))]~.
\eeq
The interference term consists of two parts: antisymmetric (the first term in Eq.~\eqref{eqn:sigmaint}) and symmetric (the second term in Eq.~\eqref{eqn:sigmaint}) parts around
Higgs boson's mass.
It is worthy to notice that at leading order
 $\oIm\left(c^{\text{SM}}_g c^{\text{SM}}_\gamma \right)$
is suppressed by $m_b/m_t$ compared to $\oRe\left(c^{\text{SM}}_g c^{\text{SM}}_\gamma \right )$
because the imaginary parts of $c^{\text{SM}}_g$, $c^{\text{SM}}_\gamma$ are mainly
 from bottom quark loop while their real parts are from
top quark or $W$ boson loop.
Thus the symmetric part of the interference term is much suppressed at leading order
and its integral value contributed for the total cross section is mainly from Next-to-Leading order~\cite{Dicus:1987fk,Campbell:2017rke}.
By contrast, the antisymmetric part could have a larger magnitude around $M_H$.

The observable $A_{\text{int}}$
 extracts the
antisymmetric part of the interference
by an sign-reversed integral around $M_H$, which is defined as

\beq
A_{\text{int}}(\xi_\gamma)=\frac{\int dM_{\gamma\gamma}
\frac{d\sigma_{\text{int}}}{dM_{\gamma\gamma}}
\Theta(M_{\gamma\gamma}-M_H)}
{\int dM_{\gamma\gamma}\frac{d\sigma_{\text{sig}}
}{dM_{\gamma\gamma}}
 }~,
\label{eqn:aint_definition}
\eeq
where the integral region is around Higgs resonance
(e.g. $[121,131]$~GeV for $M_H=126$~GeV),
 the $\Theta$-function is

\begin{center}
\begin{math}
\Theta(x)\equiv
\left\{
\begin{array}{rc}
-1, & x<0 \\
1, & x>0
\end{array}
\right.~.
\end{math}
\end{center}
So the numerator keeps the antisymmetric contribution from the interference, and the
denomenator is the cross section
from the signal, $A_{\text{int}}$ is an observable
that roughly indicates the ratio of the interference to the signal.

As $\xi_\gamma=0$ represents the SM case, we could define
 $A_{\text{int}}^{\text{SM}}\equiv A_{\text{int}}(\xi_\gamma=0)$ and
 rewrite $A_{\text{int}}(\xi_\gamma)$ simply as

\beq
 A_{\text{int}}(\xi_\gamma) = A_{\text{int}}^{\text{SM}}\times \cos\xi_\gamma ~.
\label{eqn:aint_bsm}
\eeq

The largest deviation $A_{\text{int}}(\pi) = -A_{\text{int}}^{\text{SM}}$ happens when $\xi_\gamma = \pi$, which represents an inverse CP-even $H\gamma\gamma$ coupling
from new physics but without CP violation.
It's interesting that this degenerate coupling could only be exhibited by the interference effect.

\section{Numerical Results\label{section:simulation}}

The numerical results are obtained for proton-proton collision with $\sqrt{s}=14$~TeV by using the \texttt{MCFM}~\cite{Campbell:2011bn} package, in which the subroutines according to the helicity amplitudes of Eq.~\eqref{eqn:amp} are added.
The Higgs boson's mass and width are set as $M_H=126$~GeV, and $\Gamma_H=4.3$~MeV.
Each photon is required to have $p^{\gamma}_T>20$~GeV and $|\eta^{\gamma}|<2.5$.
Based on the simulation, we study $A^{\text{SM}}_{\text{int}}$ firstly and then
$A_{\text{int}}$ from $CP$ violation cases.
After that, we estimate the feasibility to extract $A_{\text{int}}$ out at LHC.

\subsection {$A_{\text{int}}^{\text{SM}}$}

\begin{figure}[htbp]
\begin{center}
\includegraphics[width=0.5\textwidth]{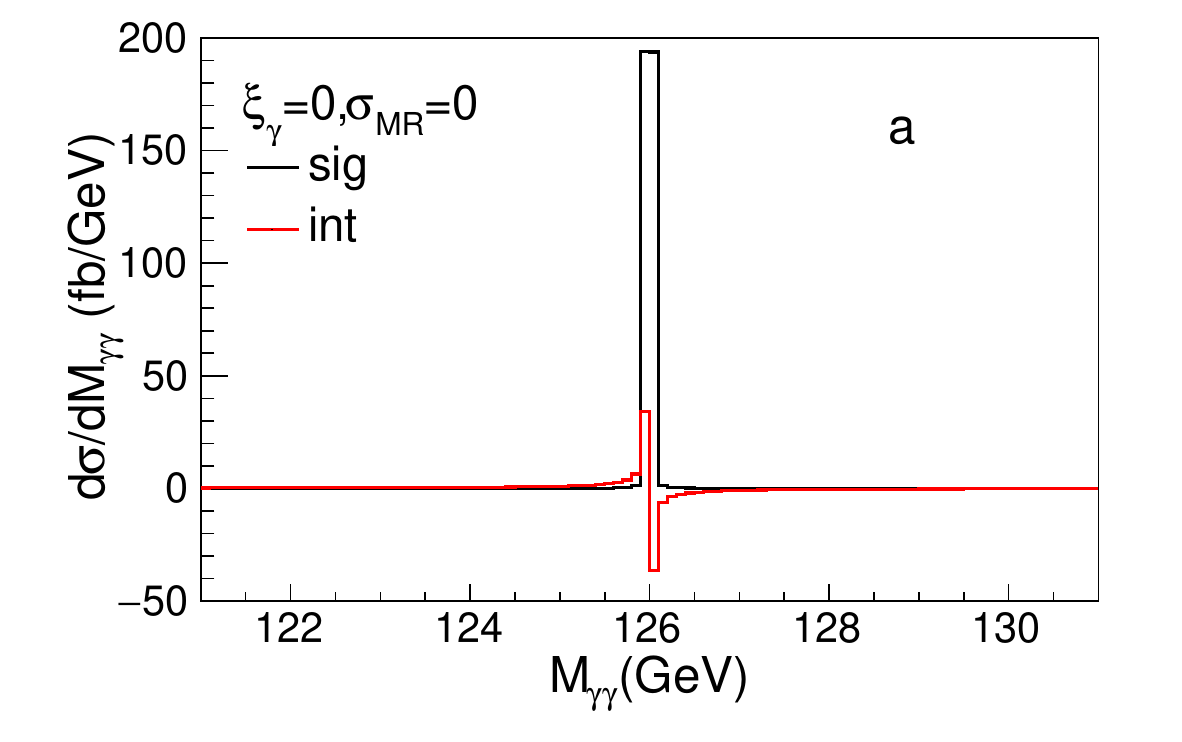}
\includegraphics[width=0.5\textwidth]{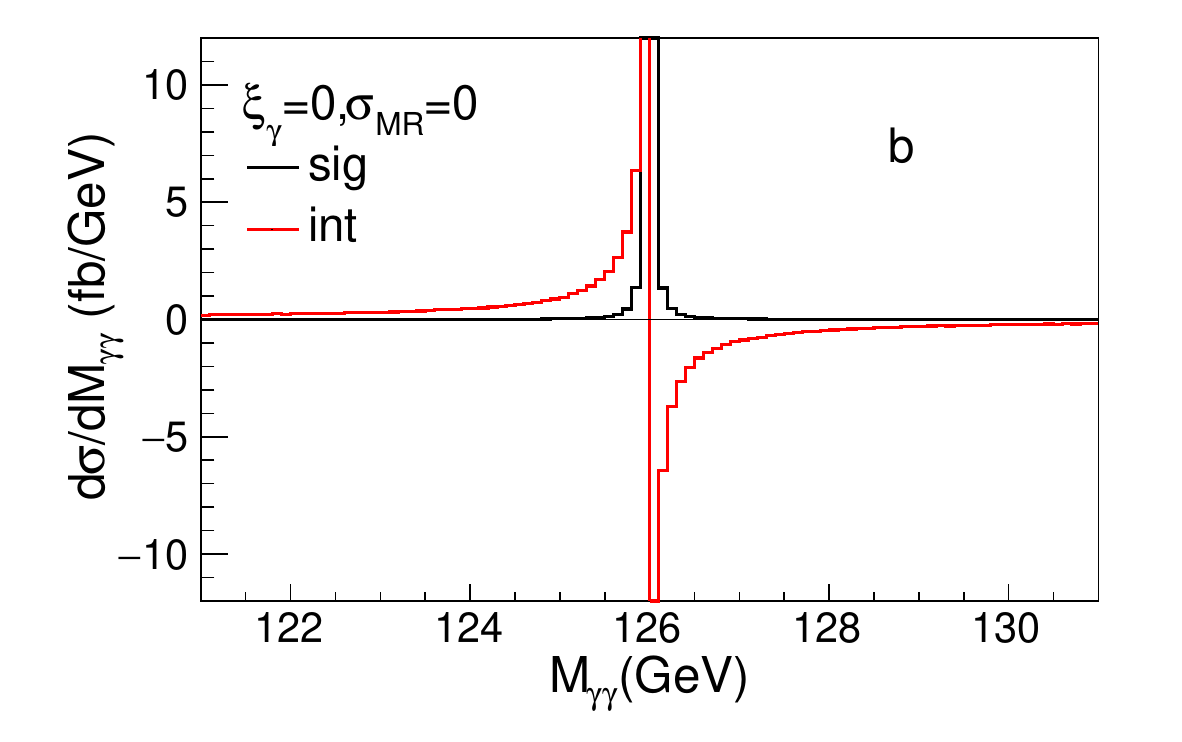}
\includegraphics[width=0.5\textwidth]{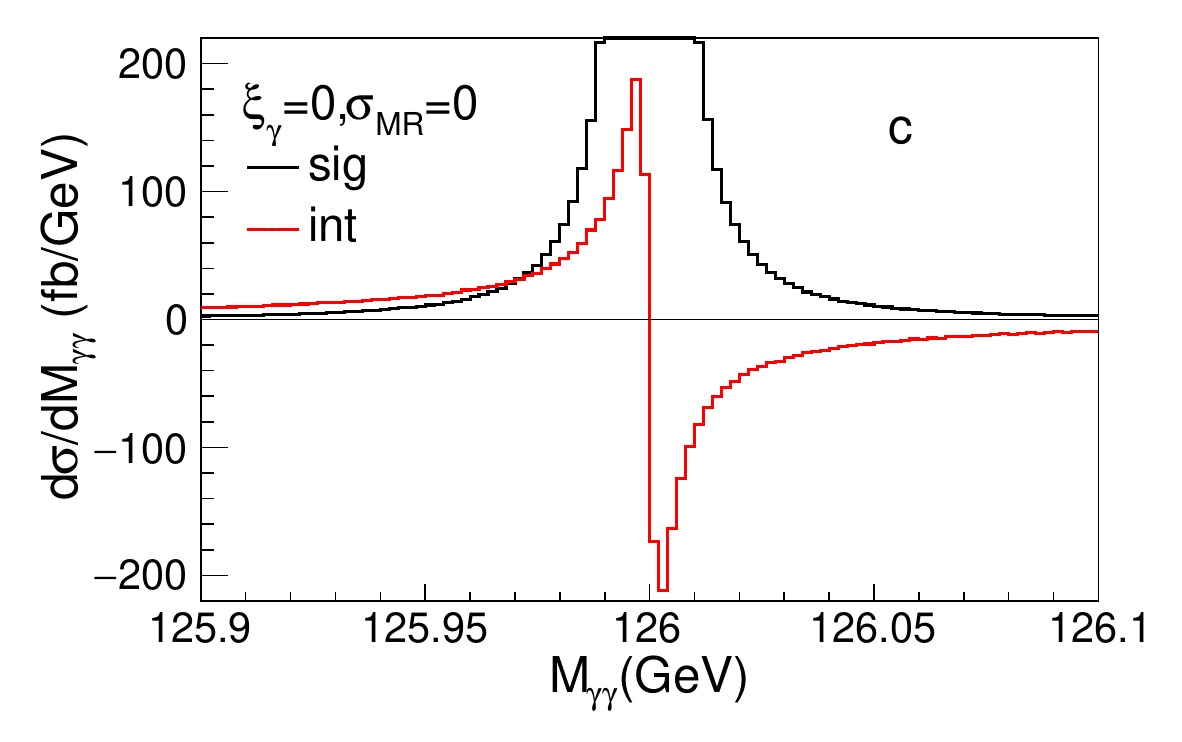}
\end{center}
\caption{ The diphoton invariant mass $M_{\gamma\gamma}$ distribution of
the signal and the interference as in
Eq.~\eqref{eqn:sigmasig} and~\eqref{eqn:sigmaint}. $\xi_\gamma=0$
represents the SM case, $\sigma_{\text{MR}}=0$ represents
the theoretical distribution before Gaussian smearing.
Among them (a) is an overall plot, (b) and (c) are close-ups.
 }
\label{fig:beforesmear}
\end{figure}

Fig.~\ref{fig:beforesmear} show
the theoretical lineshapes of the signal (a sharp peak shown in the black histogram) and the interference (a peak and dig shown in the red histogram), among which
Fig.~\ref{fig:beforesmear}a is an overall plot, Fig.~\ref{fig:beforesmear}b and Fig.~\ref{fig:beforesmear}c are close-ups.
As shown in Fig.~\ref{fig:beforesmear}a and Fig.~\ref{fig:beforesmear}b, the signal has a mass peak that is about four times of the interference, but the mass peak of the interference is wider and has a much longer tail.
The resonance region $[125.9, 126.1]$GeV is further scrutinized in Fig.~\ref{fig:beforesmear}c with bin width changed from 100~MeV to 2~MeV.
 The value of the signal exceeds that of the interference at the energy point
 $M_{\gamma\gamma}\approx M_H-10\times\Gamma_H$. 
 After integrating, the $A_{\text{int}}^{\text{SM}}$ is $~36\%$ as
shown in table~\ref{tab:aism}, which is quite large.
 As the smearing by the mass resolution (MR) is not considered yet,
 we mark it as the $\sigma_{\text{MR}}=0$ case.

\begin{figure}[htbp]
\begin{center}
\includegraphics[width=0.5\textwidth]{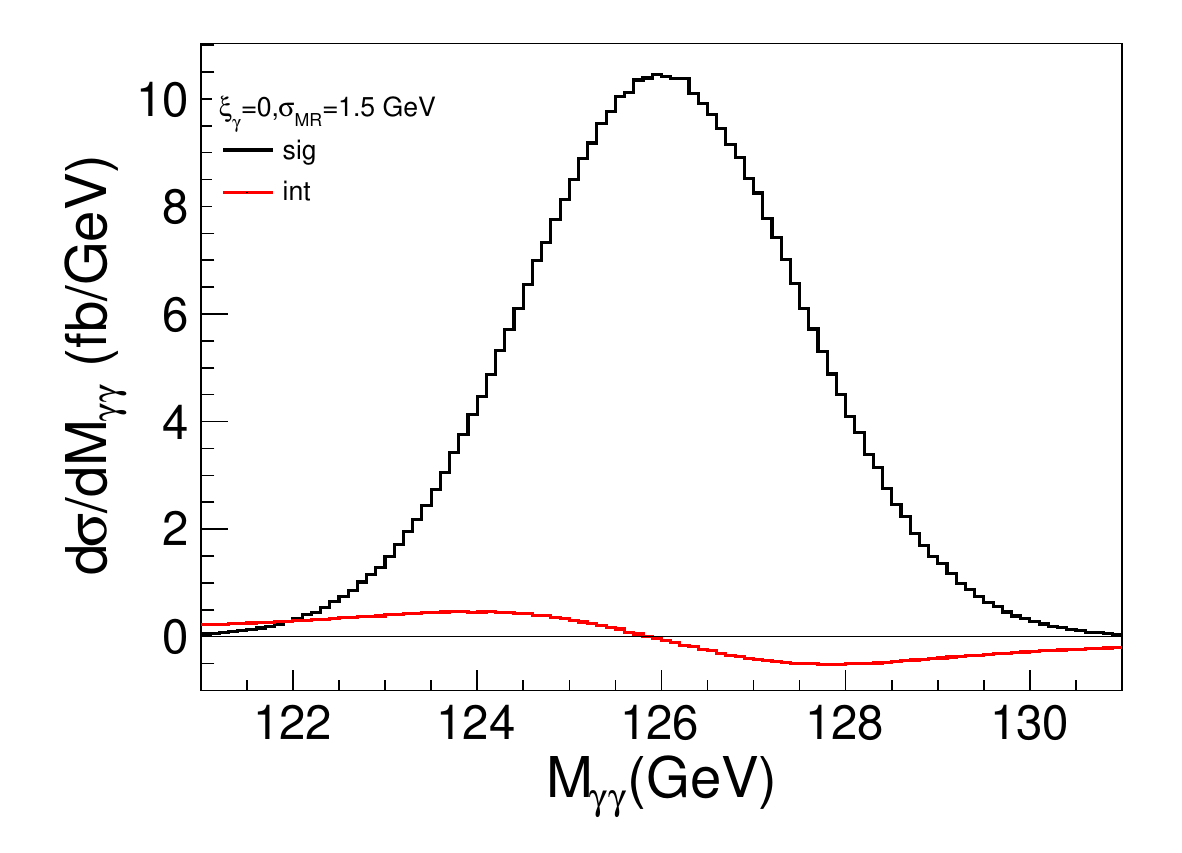}
\end{center}
\caption{ The diphoton invariant mass $M_{\gamma\gamma}$ distribution after Gaussian smearing with its mass resolution
width $\sigma_{\text{MR}}=1.5$~\text{GeV}.  }
\label{fig:aftersmear}
\end{figure}

The invariant mass of the diphoton $M_{\gamma\gamma}$ has a mass resolution
of about $1\sim2$~GeV at the LHC experiment~\cite{CMS:2017rli}.
For simplicity we include the mass resolution effect by convoluting
the histograms with a Gaussian function of width
$\sigma_{\text{MR}}=1.1, 1.3, 1.5, 1.7, 1.9$~GeV.
This convolution procedure is also called Gaussian smearing.
Fig.~\ref{fig:aftersmear} shows the lineshapes after the Gaussian smearing
with $\sigma_{\text{MR}}=1.5$~GeV.
The sharp peak of the signal becomes a wide bump (the black histogram),
meanwhile, the peak and dig of the interference are also widened,
but they cancel each other a lot near $M_H$ and
the former
peak and dig becomes a moderate antisymmetric shape around $M_H$.  \black
(the red histogram).
The $A_{\text{int}}^{\text{SM}}$ after Gaussian smearing is thus much reduced,
which range from $10.2\%$ to $7.2\%$ when $\sigma_{\text{MR}}$ goes from $1.1$ to $1.9$~GeV
as shown in table~\ref{tab:aism}.

\begin{table}[tbp]
\begin{center}
\caption{ The $A^{\text{SM}}_{\text{int}}$ values with different mass resolution widths.
The $\sigma_{\text{MR}}=0$ represents the theoretical case before Gaussian smearing.}
\label{tab:aism}
\begin{tabular}{c|c|c|c}
\hline
$\sigma_{\text{MR}}$ & $A^{\text{SM}}_{\text{int}}$ denominator & $A^{\text{SM}}_{\text{int}}$ numerator & $A^{\text{SM}}_{\text{int}}$ \tabularnewline
(GeV)& (fb)  &(fb)  & $(\%)$ \tabularnewline
 \hline
 0  & $39.3$  & $14.3$ &  $36.3$   \tabularnewline
 \hline
1.1 & $39.3$ & $4.0$  &  $10.2$    \tabularnewline
1.3 & $39.3$ & $3.7$  &  $9.4$    \tabularnewline
1.5 & $39.3$ & $3.4$  &  $8.6$    \tabularnewline
1.7 & $39.3$ & $3.1$  &  $7.9$    \tabularnewline
1.9 & $39.3$ & $2.8$  &  $7.2$    \tabularnewline
\hline
\end{tabular}
\end{center}
\end{table}

\subsection{$A_{\text{int}}(\xi_\gamma\ne 0)$}

\begin{figure}[tbp]
\begin{center}
\includegraphics[width=0.5\textwidth]{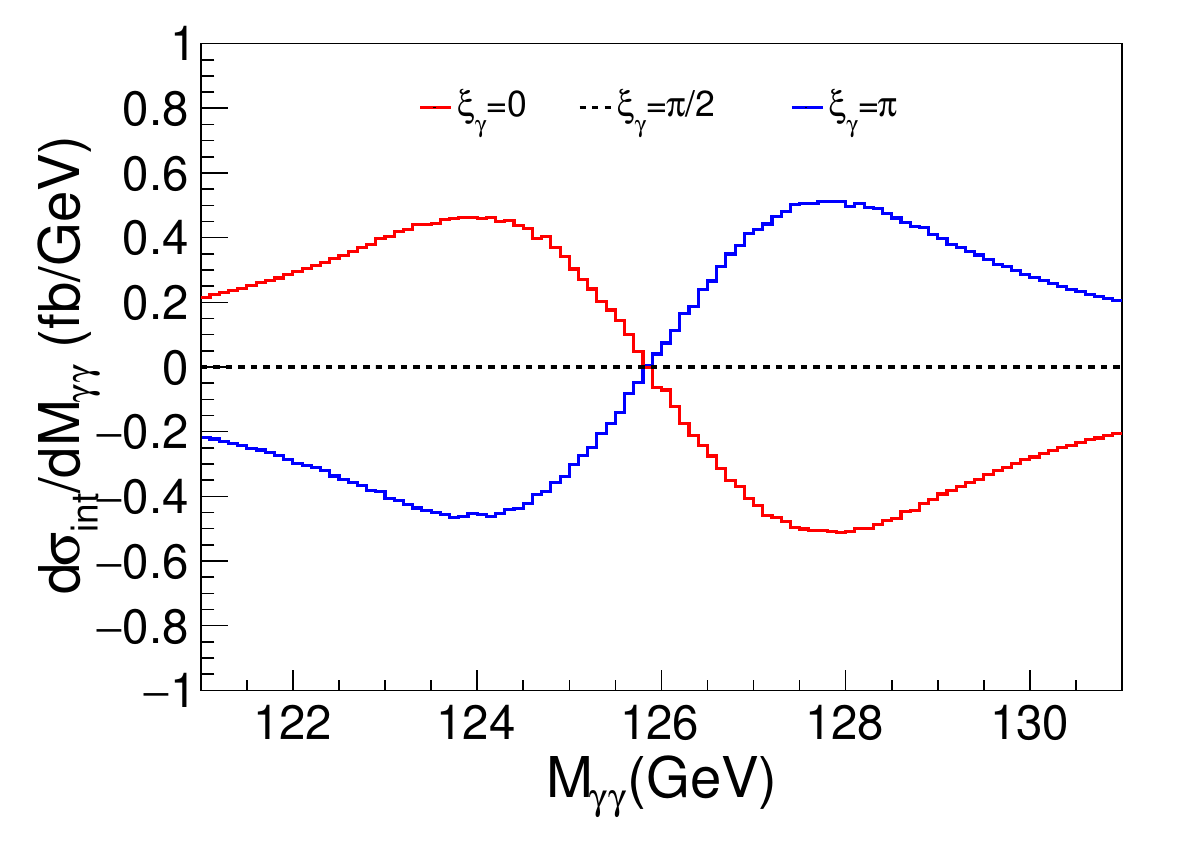}
\end{center}
\caption{ The diphoton invariant mass $M_{\gamma\gamma}$ distribution
of interference after Gaussian smearing with $\sigma_{\text{MR}}=1.5$~GeV
 when $\xi_\gamma=0,~\pi,~\pi/2$.
}
\label{fig:xigamma}
\end{figure}

\begin{figure}[tbp]
\begin{center}
\includegraphics[width=0.5\textwidth]{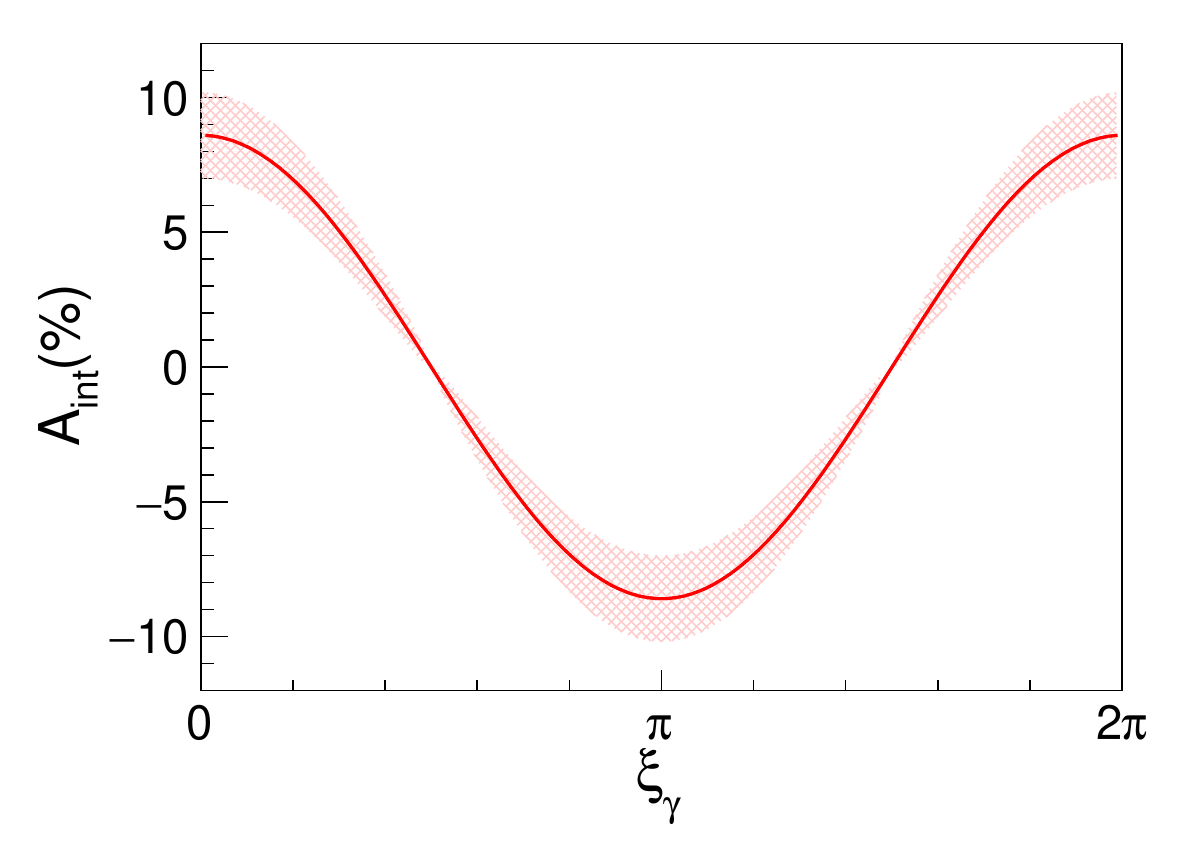}
\end{center}
\caption{ $A_{\text{int}}$ values (red line) and its statistical error (shade)
with different phase $\xi_\gamma$.
}
\label{fig:aint}
\end{figure}

Fig.~\ref{fig:xigamma} shows the lineshapes of interference under
the $\xi_\gamma=0, \pi, \pi/2$ cases with $\sigma_{\text{MR}}=1.5$~GeV.
 The blue histogram ($\xi_\gamma=\pi$, sign-reversed $CP$-even $H\gamma\gamma$ coupling) is
almost opposite to the red histogram ($\xi_\gamma= 0$, SM), which correspond to
the minimum and the maximum of $A_{\text{int}}$ values.
The black dashed
histogram ($\xi_\gamma=\pi/2$, $CP$-odd $H\gamma\gamma$ coupling) looks like a flat line
(actually with some tiny fluctuation from simulation), and it corresponds to zero $A_{\text{int}}$ value.
Fig.~\ref{fig:aint} shows $A_{\text{int}}$ and its absolute statistical error $\delta A_{\text{int}}$.
The statistical error is estimated with an integrated luminosity of 30~fb$^{-1}$, and the efficiency of detector is assumed to be one.
$\delta A_{\text{int}}$ decrease as
$A_{\text{int}}$ becomes smaller,
however, the relative statistical error $\delta A_{\text{int}}/A_{\text{int}}$
increase quickly and becomes very large as $A_{\text{int}}$ approaches zero.
In SM ($\xi_\gamma= 0$ in Fig.~\ref{fig:aint}), the relative statistical error $\delta A_{\text{int}}/A_{\text{int}}$
 is about $18\%$ with an assumption of zero correlation between symmetric and antisymmetric cross-sections.

\subsection{$A_{\text{int}}$ at LHC}

In current CMS or ATLAS experiment, the $\gamma\gamma$ mass spectrum is fitted
by a signal function and a background function. To consider the interference effect, the antisymmetric lineshape should also be included.
That is, instead of
a Gaussian function (or a double-sided Crystal Ball function) as the signal function in current LHC experiment~\cite{CMS:2017rli,Aaboud:2018xdt},
 a Gaussian function (or a double-sided Crystal Ball function) plus an asymmetric function should be used as the modified signal function, while the background function is kept as same as in the experiment.

\begin{figure}[htbp]
\begin{center}
\includegraphics[width=0.5\textwidth]{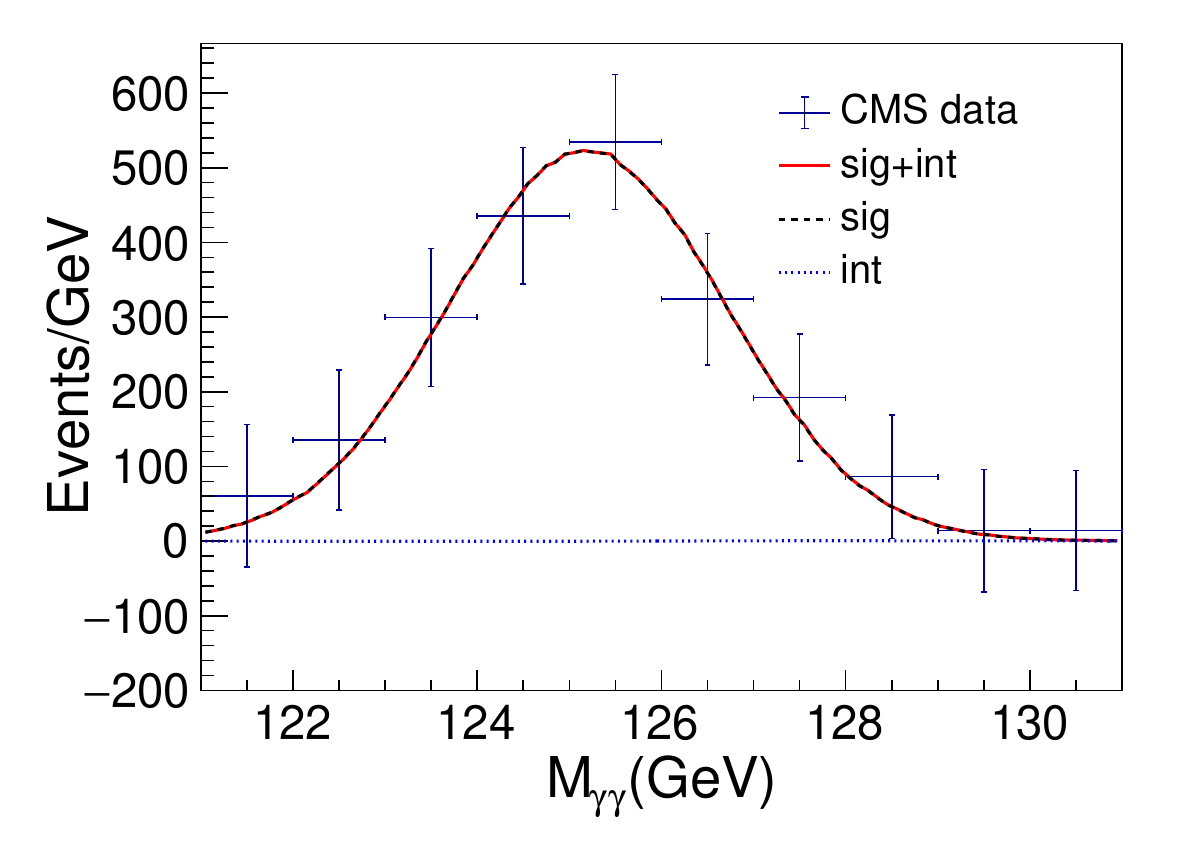}
\end{center}
\caption{A fitting on the background-substracted CMS data sample.
The crosses represent CMS data from Ref.~\cite{CMS:2017rli}.
The red solid line is the combined function, the black dashed line and the blue dotted line represent the signal and interference components respectively.
}
\label{fig:cms}
\end{figure}

\begin{figure}[htbp]
\begin{center}
\includegraphics[width=0.5\textwidth]{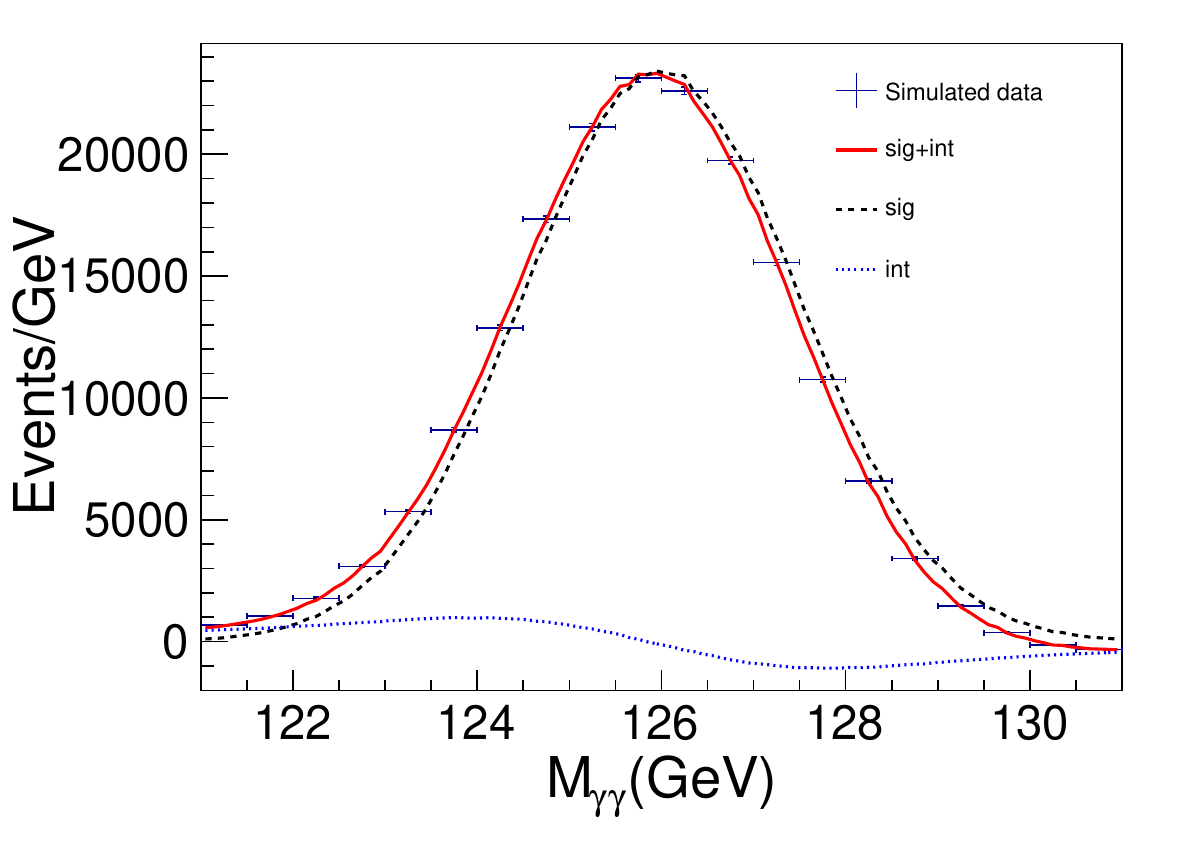}
\end{center}
\caption{
A fitting on the simulated data sample.
The crosses represent simulated data from the combined histogram in Fig.~\ref{fig:aftersmear} normalized to a luminosity of 3000~fb$^{-1}$ .
The red solid line is the combined function, the black dashed line and the blue dotted line represent the signal and interference components respectively.
  }
\label{fig:pseudo}
\end{figure}

To see whether or not the asymmetric lineshape could be extracted,
we carry out a modified-signal fitting on two background-subtracted
 data samples.
As the background fluctuation would be 
dealt with \black similarly
as in real experiment, we ignore it here for simplicity.
One data sample is from the CMS experiment in Ref.~\cite{CMS:2017rli}, we fetch 10 data points with its errors between $[121,131]$~GeV
 in the background-subtracted $\gamma\gamma$ mass spectrum
 for $35.9$~fb$^{-1}$ luminosity with proton-proton
collide energy at 13~TeV (see Fig.~13 in Ref.~\cite{CMS:2017rli}).
The fitting function is described as
\beq
f(m)=c_1\times f_{\text{sig}}(m-\delta m)+c_2\times f_{\text{int}}(m-\delta m),
\eeq
where $c_1, c_2, \delta m$ are float parameters, $m$ means the value of
 the $\gamma\gamma$ invariant mass, the functions $f_{\text{sig}}(m), f_{\text{int}}(m)$
are evaluated from the two histograms in Fig.~\ref{fig:aftersmear}
and they describe the signal and interference separately.
Fig.~\ref{fig:cms} shows the fitting result on the CMS data, in which the crosses represent CMS data with its error,
 the red solid line is the combined function, the black dashed line and the blue dotted line represent the signal and interference components respectively.
The black dashed line is almost same as the red solid line while the blue dotted line is almost flat,
 the fitting parameter $c_2$ for the interference component
 has a huge uncertainty 
that is even \black larger than the central value of $c_1$, which indicates the interference component is
hard to be extracted from
the 35.9~fb$^{-1}$ CMS data.
For a comparison, we simulate a pseudodata sample from the combined histogram in Fig.~\ref{fig:aftersmear},
 which is normalized to have events of about 80 times the CMS data (corresponding to a luminosity of 3000~fb$^{-1}$), with a binwidth of $0.5$~GeV and Poission fluctuation. The fitting result is shown in Fig.~\ref{fig:pseudo},
where the red solid line has a shift from the black dashed line,
the blue dotted line could be distinguishable clearly.
$c_1$ and $c_2$ are fitted as $c_1= 0.999\pm 0.002 $ , $c_2=0.947 \pm 0.028 $, which are consistent with their SM expected value 1 and deduce to a relative error of $A_{int}\sim 3\%$ according to the error propagation formula.
\black
Even though this fitting result looks quite good, it can only reflect
that the antisymmetric lineshape could be extracted out when no
contamination comes from systematic error.
Furthermore, our study shows that the optimal fitting strategy is taking Higgs mass as a free parameter together with $c_1$ and $c_2$. Although $M_H$ has been measured in many channels, its fluctuation is usually too large to get a converged fitting if we take it as a known input value.
\black

By contrast,
a simulation that also study the interference effect has been carried out
with systematic error included
by ATLAS collaboration at HL-LHC with a luminosity of 3000~fb$^{-1}$~\cite{ATL-PHYS-PUB-2013-014}.
In that simulation, the mass shift of Higgs boson caused by the interference effect
has been studied under different Higgs' width assumptions.
A pseudo-data is produced by smearing a Breit-Wigner with the resolution model
and the interference effect are described by the shift of smeared Breit-Wigner distributions.
Based on fitting, the mass shift of Higgs from the interference effect is estimated to be $\Delta m_H=-54.4$~MeV for the SM case,
and the systematic error on the mass difference is about $100$~MeV.
If using this result to estimate the mass shift effect for the non-SM $\xi_\gamma\ne 0$ cases, that would be,
$\xi_\gamma=\pi/2$ corresponds to a zero mass shift,
and $\xi_\gamma=\pi$ corresponds to a reverse mass shift of $\Delta m_H=+54.4$~MeV as shown in Fig.~\ref{fig:xigamma}.
Then the largest deviation of the mass shift from the SM case is $2\times 54.4$~MeV (when $\xi_\gamma=\pi$), which is almost covered by the systematic error of $100$~MeV.
Therefore, the non-SM $\xi_\gamma\ne 0$ cases could not be distinguished
 through this mass shift
effect.
Nevertheless, it is worthing to note that the antisymmetric lineshape of the interference
effect by theoretical calculation is quite different from the shift of two smeared Breit-Wigner distributions in ATLAS's simulation~\cite{ATL-PHYS-PUB-2013-014},
especially at the region far from the Higgs' peak, the antisymmetric lineshape of the interference
effect has a longer
 flat tail while the Breit-Wigner distribution falls fast.
The authors from ATLAS collaboration has also noticed this difference and planned to
add it to their next research~\cite{ATL-PHYS-PUB-2013-014}.

\section{CP violation in $Hgg$ coupling\label{section:hgg}}

In the above study the $Hgg$ coupling is supposed to be SM-like, furthermore,
the observable $A_{\text{int}}$ could also be used to probe $CP$ violation in $Hgg$ coupling.
In this section, we add one more parameter $\xi_g$ to describe $CP$ violation
in $Hgg$ coupling, and study $A_{\text{int}}$ following the same procedure as above.

Based on Eq.~\eqref{eqn:HEF}, one more parameter $\xi_g$ to describe $CP$ violation
in $Hgg$ coupling is added,
 and the effective Lagrangian is
modified as
\bea
        \mathcal{L}_{\rm h} &=&
\frac{c_\gamma \cos\xi_\gamma}{v}~h\,F_{\mu\nu} F^{\mu\nu} + \frac{c_\gamma \sin\xi_\gamma}{2v}~h\,F_{\mu\nu}\tilde{F}^{\mu\nu} \nonumber \\
&+& \frac{c_g \cos\xi_g}{v}~h\,G^a_{\mu\nu}G^{a\mu\nu} +
\frac{c_g \sin\xi_g}{2v}~h\,G^a_{\mu\nu}\tilde{G}^{a\mu\nu}~.
\label{eqn:HEFgg}
\eea

After that, the helicity amplitude in Eq.~\eqref{eqn:amp} and differential cross section
of interference in Eq.~\eqref{eqn:sigmaint} should be changed
correspondingly, which are

\bea
&\mathcal{M}
=
 -e^{-i h_1 \xi_g}e^{-i h_3\xi_\gamma}\delta^{h_1h_2}\delta^{h_3h_4}\delta^{a b}\frac{M^4_{\gamma\gamma}}{v^2}
\frac{4c_gc_\gamma}{M^2_{\gamma\gamma}-M^2_H+iM_H\Gamma_H}  \nonumber \\
&+ 4\alpha\alpha_s\delta^{a b}\sum_{f=u,d,c,s,b} Q^2_f
\mathcal{A}_{\text{box}}^{h_1 h_2 h_3h_4}~,
\label{eqn:ampgg}
\eea

\bea
&&\frac{d\sigma_{\text{int}}}{dM_{\gamma\gamma}}
\propto
\frac{(M^2_{\gamma\gamma}-M^2_H)\oRe\left(c_g c_\gamma \right)
+M_H\Gamma_H \oIm\left(c_g c_\gamma \right)
}
{(M^2_{\gamma\gamma}-M^2_H)^2+M^2_H\Gamma^2_H}
\nonumber\\
&&\times
\int dz[\cos(\xi_g+\xi_\gamma)\mathcal{A}_{\text{box}}^{++++}+\cos(\xi_g-\xi_\gamma)\mathcal{A}_{\text{box}}^{++--}]~.
\label{eqn:sigmaintgg}
\eea

Then $A_{\text{int}}^{\text{SM}}\equiv A_{\text{int}}(\xi_g=0,\xi_\gamma=0)$ and
\bea
&& A_{\text{int}}(\xi_g,\xi_\gamma) = A_{\text{int}}^{\text{SM}}\times
\nonumber \\
&& \frac{\int dz[\cos(\xi_g+\xi_\gamma)\mathcal{A}_{\text{box}}^{++++}+\cos(\xi_g-\xi_\gamma)\mathcal{A}_{\text{box}}^{++--}]}{\int dz[\mathcal{A}_{\text{box}}^{++++}+\mathcal{A}_{\text{box}}^{++--}]}~,
\label{eqn:aint_bsmgg}
\eea
where the integral could be calculated numerically once the
the integral region of $z$ is given. For example,
if the pseudorapidity of $\gamma$ is required to be $|\eta^{\gamma}|<2.5$,
that is, $z\in [-0.985,0.985]$,
the integral $\int dz \mathcal{A}_{\text{box}}^{++--}\approx -9$,
and Eq.~\eqref{eqn:aint_bsmgg} could be simplified as
\beq
A_{\text{int}}(\xi_g,\xi_\gamma)\approx A_{\text{int}}^{\text{SM}}\times
\frac{2\cos(\xi_g+\xi_\gamma)-9\cos(\xi_g-\xi_\gamma)}
{-7}~.
\label{eqn:aint_bsmgg2}
\eeq
$A_{\text{int}}(\xi_g,\xi_\gamma)$ thus has a maximum and minimum
of about $1.6$ times of $A_{\text{int}}^{\text{SM}}$.
If $\xi_g=0$, $A_{\text{int}}(\xi_g=0,\xi_\gamma)$ will degenerate to the
$A_{\text{int}}(\xi_\gamma)$ in Eq.~\eqref{eqn:aint_bsm}.
By constrast, if $\xi_\gamma=0$,
\beq
A_{\text{int}}(\xi_g)= A_{\text{int}}^{\text{SM}}\times \cos(\xi_g),
\eeq
which shows the same dependence of $A_{\text{int}}(\xi_\gamma)$
on $\xi_\gamma$ when $\xi_g=0$ as in Eq.~\eqref{eqn:aint_bsm}.
So a $CP$-violating $Hgg$ coupling could cause similar deviation of
$A_{\text{int}}$ to $A_{\text{int}}^{\text{SM}}$ as a $CP$-violating $H\gamma\gamma$ coupling, and
an single observed $A_{\text{int}}$ value could not distinguish them
 since there are two free parameters for
one observable.

\begin{figure}[tbp]
\begin{center}
\includegraphics[width=0.5\textwidth]{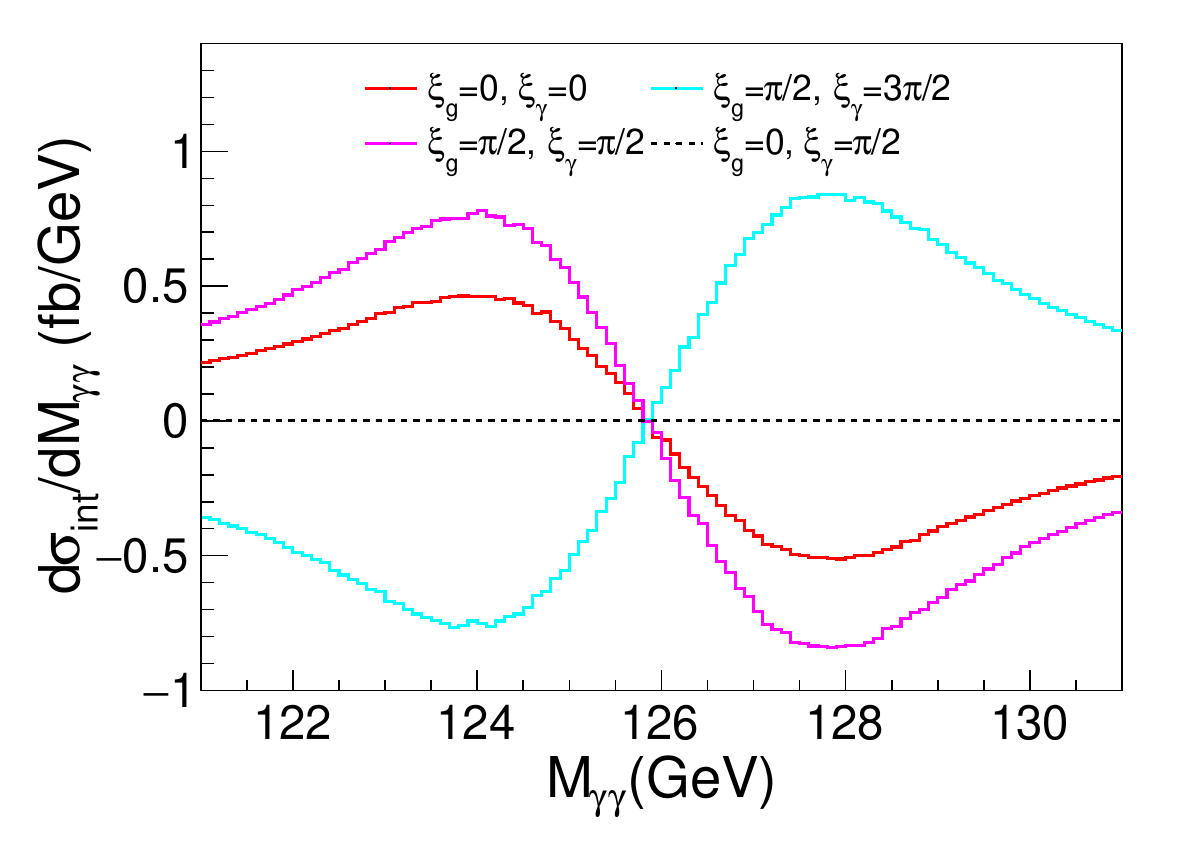}
\end{center}
\caption{ The diphoton invariant mass $M_{\gamma\gamma}$ distribution
of interference after Gaussian smearing
in various $\xi_g,~\xi_\gamma$ cases. }
\label{fig:xigxigamma}
\end{figure}

Fig.~\ref{fig:xigxigamma} shows the lineshapes of interference for different
$\xi_g,~\xi_\gamma$ choices.
The red histogram ($\xi_g=0,~~\xi_\gamma=0$) represents the SM case; the magenta histogram
($\xi_g=\frac{\pi}{2},~~\xi_\gamma=\frac{\pi}{2}$) could get largest $A_{\text{int}}$;
the cyan histogram ($\xi_g=\frac{\pi}{2},~~\xi_\gamma=\frac{3\pi}{2}$) corresponds
 to the smallest $A_{\text{int}}$ ; and the black histogram
is from $\xi_g=0,~~\xi_\gamma=\frac{\pi}{2}$ case with $A_{\text{int}}$ equal to zero.
 For the general case of both $\xi_g,~\xi_\gamma$ being free parameters,
$A_{\text{int}}(\xi_g,~\xi_\gamma)$ could have a wider value range than $A_{\text{int}}(\xi_\gamma)$, which makes it easier to be probed in future experiment.

\section{conclusion and discussion\label{section:conclusion}}

The diphoton mass distribution from
the interference between $gg\to H \to \gamma\gamma$ and $gg\to \gamma\gamma$
at leading order is almost antisymmetric around $M_H$ and we propose
an sign-reversed integral around $M_H$ to get its contribution.
After dividing this integral value by the cross section of Higgs signal,
we get an observable $A_{\text{int}}$.
In SM, the theoretical $A_{\text{int}}$ value 
 before taking into account
the mass resolution could be $\sim39\%$. \black
After considering mass resolution
of $\sim1.5$~GeV, $A_{\text{int}}$ is reduced but still
could be as large as $10\%$.
The $CP$ violation in $H\gamma\gamma$ could change $A_{\text{int}}$ from
$10\%$ to -$10\%$ depending on the $CP$ violation phase $\xi_\gamma$.
In a general framework of both $CP$-violating $H\gamma\gamma$ and $Hgg$ coupling,
$A_{\text{int}}$ could have a larger value of $\sim \pm 16\%$.
However, due to the systematic error and statistical error are both $\sim 10\%$ in
current experiments at LHC, the antisymmetric lineshape is difficult to be extracted out.
Even at futuristic high luminosities, the large systematic error is still a tricky obstacle.

\begin{acknowledgments}
The work is supported by the National Natural Science Foundation of China under Grant No.11405102 and No.11847168, and the Fundamental Research Funds for the Central Universities of China under Grant No. GK201603027 and No. GK201803019.

\end{acknowledgments}

\appendix

\bibliographystyle{utphys}
\bibliography{reference}

\end{document}